\documentclass[11pt,fleqn]{article}

\usepackage{latexsym}

\newcommand{\be}{\begin{equation} } 
\newcommand{\ee}{\end{equation} \par \noindent}

\newcommand{\rf}[1]{(\ref{#1})}

\newcommand{\scal}[2]{\mbox{$\langle #1 \! \mid #2 \rangle $}} 

\newcommand{\ba}{\begin{array}}
\newcommand{\ea}{\end{array}}
\newcommand{\const}{{\rm const}}

\newcommand{\C}{{\bf C}}
\newcommand{\R}{{\bf R}}

\newcommand{\m}{\left( \ba{c}}
\newcommand{\ema}{\ea \right)}
\newcommand{\mm}{\left( \ba{cc}}

\newtheorem{lem}{Lemma}

\newenvironment{Proof}{\par \vspace{2ex} \par
\noindent \small {\it Proof:}}{\hfill $\Box$ 
\vspace{2ex} \par }

\begin{document}

\title{\bf 
A new  approach to the Darboux-B\"acklund \\ transformation {\it versus}
the standard dressing method}

\author{{\bf
Jan L.\ Cie\'sli\'nski\thanks{
E-mail: \tt janek\,@\,alpha.uwb.edu.pl}}
\\ {\footnotesize Uniwersytet w Bia\l ymstoku,  
Instytut Fizyki Teoretycznej}
\\ {\footnotesize ul.\ Lipowa 41, 15-424  
Bia\l ystok, Poland} 
\\ {\footnotesize \vspace{-2ex}}
\\ {\bf Waldemar Biernacki\thanks{E-mail: 
\tt wb\,@\,sao.pl}}
\\ {\footnotesize Wy\.zsza Szko\l a Ekonomiczna w Bia\l ymstoku, Katedra Informatyki}
\\ {\footnotesize ul.\ Choroszcza\'nska 31, 15-732 Bia\l ystok, Poland}
}

\maketitle

\begin{abstract} 
We present a new approach to the construction of the Darboux matrix.
This is a generalization of the recently formulated method based 
on the assumption that the square of the Darboux matrix 
vanishes for some values of the spectral parameter. 
We consider the multisoliton case, the reduction problem
and the discrete case. The relationship between 
our approach and the standard dressing
method is discussed in detail.
\end{abstract}

\par \vspace{0.5cm} \par
\par \vspace{0.5cm} \par
\par \vspace{0.5cm} \par

\section{Introduction}

There are several methods to construct the Darboux matrix 
(which generates soliton solutions)  
\cite{ZS,Mi,NM,Its,Gu,MS,Ci-dbt,RS}). However, these methods are
technically difficult when applied to the matrix versions of the spectral 
problems which are naturally represented in Clifford algebras
\cite{CGS-izot,Ci-izot,Ci-TMP}. 
Some of these problems are avoided in our recent paper 
\cite{BC}. 
In the present paper we develop the ideas of \cite{BC} in the matrix case.
We extend our approach on the multisoliton case and consider the reduction problem
and the discrete case. We also 
show that our approach, although different, is to some extent 
equivalent to the standard dressing
method. We compare our method with the Zakharov-Shabat approach \cite{ZS,ZMNP} 
and the Neugebauer-Meinel approach \cite{NM,MNS}.

We consider the spectral problem 
\be     \label{problin}
    \Psi_{,\mu} = U_\mu\Psi , \qquad (\mu=1,\ldots,m) 
\ee
(with no assumptions on $U_\mu$ except  rational dependence on $\lambda$)
and the Darboux transformation
\be   \label{DBT}
   \tilde{\Psi} = D\Psi \ ,                          
\ee
which means that
\be  \label{tilproblin}
  \tilde{\Psi},_{\mu} = \tilde{U}_\mu \tilde{\Psi} \ ,
\ee
where $\tilde{U}_\mu$ and $U_\mu$ have the same rational dependence 
on $\lambda$ ($U_\mu$ and $\Psi$ are $n \times n$ matrices but our approach works well also in the Clifford numbers case \cite{BC}). 

The construction of the Darboux transformation is
well known (especially in the matrix case) \cite{Ci-dbt,ZMNP}. 
The first step is the equation
for $D$ resulting from \rf{problin},\rf{DBT} and \rf{tilproblin}:
\be  \label{Dmu}
   D,_\mu + D U_\mu = \tilde{U}_\mu D  \ .
\ee

In our erlier paper \cite{BC} we proposed the following procedure.
We assume that 
there exist two different values of $\lambda$, say
$\lambda_+$ and $\lambda_-$, satisfying 
\be  \label{kwadrat} 
D^2 (\lambda_\pm) = 0 \ .
\ee
 Denoting
$\Psi(\lambda_\pm)=\Psi_\pm$, $D(\lambda_\pm)=D_\pm$, evaluating \rf{Dmu} at 
$\lambda = \lambda_\pm$ and multiplying \rf{Dmu} 
by $D_\pm$ from the right, we get:
\be  \label{eqforD}
  D_\pm,_\mu D_\pm + D_\pm U_\mu (\lambda_\pm) D_\pm = 0 \ .
\ee
We assume that $\Psi (\lambda_\pm)$ are invertible (which is obviously 
true in the generic case). 
It is not difficult to check that $D_\pm$ given by
\be  \label{Dpm}
   D_\pm = \varphi_\pm \Psi_\pm d_\pm \Psi_\pm^{-1} \ , \qquad   
d_\pm^2 = 0 \ ,
\ee
(where $d_\pm = \const$ and $\varphi_\pm$ are scalar functions) 
satisfy  equations \rf{kwadrat}, \rf{eqforD}. Assuming that $D$ is linear in $\lambda$, i.e.,
\be \label{Dlin}
  D (\lambda) = A_0 + A_1 \lambda \ , 
\ee
we can easily express  $A_0, A_1$ by $D_\pm$ to get
\be  \label{newD}   D(\lambda) = 
\frac{\lambda-\lambda_-}{\lambda_+ -\lambda_-} \varphi_+ \Psi_+ d_+  \Psi_+^{-1} +
\frac{\lambda-\lambda_+}{\lambda_- -\lambda_+} \varphi_- \Psi_- d_- \Psi_-^{-1}  \ .
\ee

\section{One-soliton case and the Zakharov-Shabat approach}

We confine ourselves to the  case linear in $\lambda$ (see \rf{Dlin}). 
The condition \rf{kwadrat} can be easily realized if 
\be  \label{Dlamu}
     D^2 (\lambda) = \sigma (\lambda-\lambda_+)(\lambda-\lambda_-) I
\ee
where $\sigma \neq 0$ is a constant, $\lambda_+ \neq \lambda_-$ and $I$ is the identity matrix. 
The identity matrix will be sometimes omitted 
(i.e.,  for $a \in \C$ we write $a I  = a$). 
In the case \rf{Dlamu} from \rf{kwadrat} and \rf{newD} it follows that 
\be   \label{kwadrat2}
  D_+  D_-  + D_- D_+ = - \sigma (\lambda_+ - \lambda_-)^2 \ .
\ee

\begin{lem}
$D$ of the form \rf{Dlin} satisfies \rf{Dlamu} if and only if $n$ is even and
\be \label{oldD}
    D = {\cal N} \left( 
\lambda - \lambda_+  + (\lambda_+ - \lambda_-) P \right)
\ee
where the matrices ${\cal N}$ and $P$ satisfy
\be  \label{PN}  P^2 = P \ , \qquad 
{\cal N}^2 = \sigma \ , \qquad 
{\cal N} P {\cal N}^{-1} = I - P \ . 
\ee
In this case the Darboux matrices \rf{newD} and \rf{oldD} are equivalent.
\end{lem}
\begin{Proof} We denote ${\cal N} := A_1$. From \rf{Dlin} we get
\[
D^2 (\lambda) = A_0^2 + (A_0 {\cal N} + {\cal N} A_0) \lambda + {\cal N}^2 \lambda^2 \ , 
\]
i.e., $D^2 (\lambda)$ is a quadratic polynomial. It is proportional to the identity matrix $I$ (compare \rf{Dlamu}) iff 
\be   \label{three}
     {\cal N}^2 = \sigma  \ , \qquad 
A_0 {\cal N} + {\cal N} A_0 = - \sigma (\lambda_+ + \lambda_-)  \ , 
\qquad A_0^2 = \sigma \lambda_+ \lambda_-  \ . 
\ee
 Multiplying the second equation by ${\cal N} A_0$ we get
\[
\sigma^2 \lambda_+ \lambda_- + ({\cal N} A_0 )^2 +
 \sigma (\lambda_+ + \lambda_-)  {\cal N} A_0 = 0 \ .
\]
Hence $({\cal N} A_0 + \sigma \lambda_+) ({\cal N} A_0 + \sigma \lambda_-) = 0$, 
and, denoting $Q := {\cal N} A_0 + \sigma \lambda_+$, we have
\[
    Q^2 = (\lambda_+ - \lambda_-) \sigma Q 
\]
which means that $Q = (\lambda_+ - \lambda_-) \sigma P$, where $P^2=P$. 
Therefore, taking into account ${\cal N}^2 = \sigma$, we get \rf{oldD}. Now,
we take into account the third equation of \rf{three}. 
First, $A_0^2 P = \sigma \lambda_+ \lambda_- P$ yields  
$\lambda_- (\lambda_+ - \lambda_-) {\cal N} P {\cal N} P = 0$. Then the
equation $A_0^2 = \sigma \lambda_+ \lambda_- $ is equivalent to 
$\lambda_+ (\lambda_+ - \lambda_-) (\sigma (I - P) - {\cal N} P {\cal N}) = 0$.
Therefore ${\cal N} P {\cal N}^{-1} = I - P$. This equality means that 
$\ker P = {\cal N}^{-1} {\rm Im} P$ which implies $\dim \ker P = 
\dim {\rm Im} P$. Thus $n$ is even which complets the proof.
\end{Proof}

The case $\lambda_+ = \lambda_-$ can be treated in a similar way
and it leads to the nilpotent case \cite{Ci-dbt}:
\[
 D = {\cal N} (\lambda - \lambda_+  + M) \ , \qquad M^2 = 0 \ , 
\quad {\cal N}^2 = \sigma \ , \quad M =  - {\cal N} M {\cal N}^{-1} \ .
\]

Our method is closely related to the standard dressing 
transformation \cite{ZS,Ci-dbt,ZMNP}. The Darboux matrix \rf{oldD} 
can be rewritten as 
\be \label{oldD'}
    D = \left( \lambda - \lambda_+ \right) {\cal N} \left( I  +
\frac{\lambda_+ - \lambda_-}{\lambda - \lambda_+} P \right) \ .
\ee
We recognize the standard one-soliton Darboux matrix in the Zakharov-Shabat 
form \cite{Ci-dbt,ZMNP}. We point out that 
usually one considers the Darboux matrix 
${\cal D} = (\lambda-\lambda_+)^{-1} D$ which is equivalent to 
$D$ given by \rf{oldD} because the multiplication of $D$ by a constant factor
leaves the equation \rf{Dmu} invariant \cite{Ci-nos}.
${\cal N}$ is known as the normalization matrix and $P$ is a projector
expressed by the background wave function: 
\be \label{kerimP}
{\rm ker} P = \Psi(\lambda_+) V_{ker} \ , \qquad {\rm im} P = \Psi (\lambda_-) V_{im} \ ,
\ee
$V_{ker}$ and $V_{im}$ are some constant   
vector spaces, $\lambda_+$ and $\lambda_-$ are constant 
complex parameters. 
The last constraint of \rf{PN} has the following interpretation.
Let ${\cal N} P {\cal N}^{-1} = I - P$. Then
\[  \ba{l}
v \in {\rm im} P \  \Leftrightarrow \quad (I - P) v = 0 
\ \Leftrightarrow \quad  P {\cal N}^{-1} v = 0 
\ \Leftrightarrow \quad  {\cal N}^{-1} v \in {\rm ker} P  \\[2ex]
v \in {\rm ker} P \  \Leftrightarrow \quad  P v = 0 
\ \Leftrightarrow \quad  P {\cal N}^{-1} v = {\cal N}^{-1} v 
\ \Leftrightarrow \quad  {\cal N}^{-1} v \in {\rm im} P  
\ea \]
Hence, $\dim {\rm im} P = \dim {\rm ker} P = d \equiv n/2$, which implies 
$\dim V_{im} = \dim V_{ker}$. In this case, given a 
projector $P$, one can always find a corresponding ${\cal N}$. Indeed,
let $v_1,\ldots,v_d$ be a basis in ${\rm im} P$ and 
$w_k := {\cal N}^{-1} v_k$ ($k=1,\ldots,d$) an associated basis in 
${\rm ker} P$.
By virue of ${\cal N}^2 = \sigma$ we have ${\cal N}^{-1} w_k = \sigma^{-1} v_k$.
Therefore
\[
   {\cal N}^{-1} (v_1,\ldots,v_d,w_1,\ldots,w_d) = 
(w_1,\ldots,w_d,v_1/\sigma,\ldots,v_d/\sigma)
\] 
(where $(v_1,v_2,\ldots)$ denotes the matrix with columns $v_1,v_2,\ldots$)
and, finally, 
\be  \label{Nvw}
{\cal N} =  (v_1,\ldots,v_d,w_1,\ldots,w_d) 
(w_1,\ldots,w_d,v_1/\sigma,\ldots,v_d/\sigma)^{-1} \ .
\ee
The ${\cal N}$ obtained in this way depends on the choice of the bases
$v_1,\ldots,v_d$ and $w_1,\ldots,w_d$ (we can put $A v_k$, $\det A \neq 0$, in the place of $v_k$ and $B w_j$, $\det B \neq 0$, in the place of $w_j$). In other words, ${\cal N}$ is given up to 
nondegenerate $d \times d$ matrices $A$ and $B$. 

The formulas \rf{newD} and \rf{oldD}
coincide after appropriate identification of the parameters. 
Indeed, comparing  coefficients by powers of $\lambda$ we have:
\be  \label{N} \begin{array}{l}
{\cal N} =\frac{\displaystyle \varphi_+ \Psi_+ d_+ \Psi_+^{-1} - 
\varphi_- \Psi_- d_- \Psi_-^{-1}}{\displaystyle \lambda_+ - \lambda_-} \ , \\[3ex]
{\cal N} ( - \lambda_+ + (\lambda_+ - \lambda_-)  P ) =
\frac{\displaystyle \lambda_+ \varphi_- \Psi_- d_- \Psi_-^{-1} - 
\lambda_- \varphi_+
\Psi_+ d_+ \Psi_+^{-1}}{\displaystyle \lambda_+ - \lambda_-} \ ,
\end{array} \ee
and after straightforward computation we get
\be \begin{array}{l}   \label{P-I-P}
P = (\varphi_+ \Psi_+ d_+ \Psi_+^{-1} - \varphi_- \Psi_- d_- \Psi_-^{-1})^{-1} 
\varphi_+ \Psi_+ d_+ \Psi_+^{-1} \ , \\[3ex]
I - P = (\varphi_- \Psi_- d_- \Psi_-^{-1} - \varphi_+ \Psi_+ d_+ \Psi_+^{-1})^{-1} 
\varphi_- \Psi_- d_- \Psi_-^{-1} \ .
\end{array} \ee

Taking into account the assumption \rf{kwadrat2} we have:
\be
P = \frac{D_- D_+}{D_+ D_-  +  D_- D_+} = \frac{- D_-  D_+}{\sigma (\lambda_+ - \lambda_-)^2} \ .
\ee

The above results are valid for $n\times n$ matrix linear problems.
Now, we focus on the $2\times 2$ case.
Because the elements $d_+$, $d_-$ are nilpotent ($d_\pm=0$), then there exist vectors $v_+, v_-$ such that 
\be  \label{vw}
 d_+  v_+ = 0 \ , \qquad d_- v_- = 0 \ .
\ee
Then from \rf{P-I-P} it follows immediately 
 $ P \Psi_+ v_+ = 0 $ and 
$ (I-P) \Psi_- v_- = 0 $, i.e., $\Psi_+ v_+$ span 
${\rm ker} P$ and 
$\Psi_- v_- $ span ${\rm im} P$. Hence, $v_+ \in V_{ker}$ and $v_- \in V_{im}$.

It is not difficult to check that the general form of $2\times 2$ 
matrices $d_\pm$ such that $d_\pm^2 = 0$ is given by
\be    \label{dpm} 
d_\pm = \left( \begin{array}{cc} 
- a_\pm b_\pm & b_\pm^2 \\ -a_\pm^2 &  a_\pm b_\pm 
\end{array} \right) = \m b_\pm \\ a_\pm \ema \mm - a_\pm  & b_\pm \ema \ ,
\ee
where $a_\pm, b_\pm$ are complex numbers. Therefore, to satisfy \rf{vw},  we can take
\be
 v_+ = \left( \begin{array}{r} b_+ \\ a_+ \end{array} \right) \ , \qquad 
 v_- = \left( \begin{array}{r} b_- \\ a_- \end{array} \right) \ .
\ee 
We have almost unique correspondence (i.e., up to a scalar factor) 
between $v_+$ and $d_+$ and between $v_-$ and $d_-$. 

Denoting
\[
   \Psi_+ v_+ \equiv \m B_+ \\  A_+ \ema \ , \qquad
 \Psi_- v_- \equiv \m B_- \\  A_- \ema \ , 
\]
we get the explicit formula for $P$
\be  \label{exP}
   P  =  \mm 0 & B_- \\ 0 &  A_- \ema 
\mm B_+  & B_- \\  A_+ &   A_- \ema^{-1} = 
\frac{\mm - A_+ B_- & B_+ B_- \\
- A_+ A_- & B_+ A_- \ema }{A_- B_+ - A_+ B_- }  
 \ee
The corresponding ${\cal N}$ reads (compare \rf{Nvw}):
\be  \label{exN}
    {\cal N} = \frac{1}{A_- B_+ - A_+ B_-} 
\mm \sigma A_- B_- - A_+ B_+ & B_+^2 - \sigma B_-^2  \\ 
\sigma A_-^2 - A_+^2 &  A_+ B_+ - \sigma A_- B_- \ema
\ee
Although we can reduce our approach to the explicit formulas  
\rf{exP} and \rf{exN} the main advantage of our method consists 
in expressing the Darboux transformation in terms of 
$\Psi_\pm d_\pm \Psi_\pm^{-1}$ and avoiding difficulties with parameterizing kernel and image of the projector $P$ which is especially troublesome in
the Clifford algebras case.  

\section{Reductions}

Let us consider the unitary reduction 
\be  \label{un}
    U_\mu^\dagger (\bar \lambda) = - U_\mu (\lambda) \ .
\ee
If $U_\mu$ is a polynom in $\lambda$, then the condition 
\rf{un} means that the coefficients of this polynom by powers of $\lambda$ 
are $u (n)$-valued.

One can easily prove that \rf{un} implies 
$\Psi^\dagger (\bar \lambda)  \Psi (\lambda) = C (\lambda)$,
where $C (\lambda)$ is a constant matrix  ($C,_\nu = 0$).
The matrix $C$ can be fixed by a choice of the  
initial conditions. Usually we confine ourselves to the case 
\be  \label{Psiun}
   \Psi^\dagger (\bar \lambda)  \Psi (\lambda) = 
k (\lambda)  I  \ ,
\ee
where $k (\lambda)$ is analytic in $\lambda$. From 
\rf{Psiun} we  can derive $\overline{k ({\bar \lambda})} 
= k (\lambda)$. 
By virtue of \rf{DBT}, the Darboux matrix have to satisfy the analogical constraint:
\be \label{plam}
   D^\dagger (\bar \lambda)  D (\lambda)  = p (\lambda) I \ .
\ee
Assuming that $D$ is a polynom with respect to $\lambda$,
compare \rf{Dlin}, we get that
$p (\lambda)$ is a polynom with constant real coefficients, i.e., 
$\overline{p(\bar{\lambda})} = p (\lambda)$ and $p,_\nu = 0$.

\begin{lem}  \label{roots1}
If $D$ is linear in $\lambda$ and \rf{plam} holds, then roots of the 
equation $\det D (\lambda) = 0$ satisfy the quadratic 
equation  $p (\lambda) = 0$.
\end{lem}
\begin{Proof}
Let $p (\lambda) = \alpha \lambda^2 + \beta \lambda + \gamma$. From
\rf{Dlin}, \rf{plam} it follows
\be
   A_0^\dagger A_0 = \gamma \ , \qquad A_1^\dagger A_1 = \alpha \ , \qquad
A_0^\dagger A_1 + A_1^\dagger A_0 = \beta 
\ee
which can be easily reduced to a single equation for $S := - A_0 A_1^{-1}$.
Namely,
\be  \label{quadS}
 \alpha S^2 + \beta S + \gamma = 0 \ . 
\ee
Therefore, the eigenvalues of $S$ have to satisfy the equation $p (\lambda) = 0$.
Indeed, if $S {\vec v} = \mu {\vec v}$, then $(\alpha \mu^2 + \beta \mu + \gamma)
{\vec v} = 0$.
On the other hand, the equation $\det D(\lambda) = 0$ can be rewritten as
\be
     0 = \det (\lambda I - S) \det A_1 \ ,
\ee
which means that the roots of $\det D (\lambda) = 0$ coincide with eigenvalues of $S$.
\end{Proof}

\begin{lem}
We assume \rf{Dlamu}. Then the reduction \rf{Psiun} imposes the following 
constraints on the Darboux matrix \rf{newD}:
\be
\lambda_- = \lambda_+^\dagger \ , \qquad   d_-^\dagger  d_+ = 0 \ , 
\ee
\quad
and (for $n=2$) \ $ \scal{v_+}{v_-} = 0 $. 
\end{lem}

In particular, by virtue of \rf{kwadrat}, we can take
$d_- = f d_+^\dagger$, where $f$ is a scalar function.

\begin{Proof}  
Let us denote zeros of the polynom $p (\lambda)$ by 
$\lambda_1$, $\lambda_2$. 
Because 
$\overline{p(\bar{\lambda})} = p (\lambda)$ 
there are two possibilities: either 
$\lambda_2 = {\bar \lambda}_1$ or $\lambda_1$, $\lambda_2$ are real.
From \rf{Dlamu} we have
\be   \label{deten}
(\det D (\lambda) )^2 = 
\sigma^n (\lambda - \lambda_+)^n (\lambda - \lambda_-)^n \ .
\ee
Therefore, 
in the case \rf{Dlamu}, Lemma~\ref{roots1} means that 
$\lambda_+$, $\lambda_-$ coincide with $\lambda_1$, $\lambda_2$.

Suppose that $\lambda_+ \in \R$. Then 
from \rf{plam} we have $(D (\lambda_+) )^\dagger D (\lambda_+) = 0$
which implies $D_+ \equiv D (\lambda_+) = 0$ 
(because for any vector $v \in \C^n$
the scalar product $\scal{v}{D_+^\dagger D_+ v} = 0$, hence
$\scal{D_+ v}{D_+ v} = 0$, and, finally $D_+ v = 0$).
Therefore $\lambda_+$ (and, similarly, $\lambda_- $) cannot be real.
Thus $\lambda_- = \lambda_+^\dagger$. In this case \rf{plam} reads
\be  \label{DherD0}
 ( D (\lambda_-) )^\dagger  D (\lambda_+) = 0 \ .
\ee
Using \rf{Dpm} and \rf{Psiun} (assuming $k (\lambda_\pm) \neq 0$) we
get 
\[
   ( D (\lambda_-) )^\dagger = {\bar \varphi}_- (\Psi_-^\dagger)^{-1} 
d_-^\dagger   \Psi_-^\dagger = {\bar \varphi}_- \Psi_+ d_-^\dagger \Psi_+^{-1} 
\]
and \rf{DherD0} assumes the form
 $\varphi_+ {\bar \varphi}_-  \Psi_+ d_-^\dagger d_+ \Psi_+^{-1} = 0$.
Hence  $d_-^\dagger d_+ = 0$. 

Finally, in the case $n=2$, we use \rf{dpm}. Then the condition $d_-^\dagger d_+ = 0$ 
is equivalent to \ $a_+ {\bar a}_- + b_+ {\bar b}_- = 0$, i.e., $\scal{v_+}{v_-} = 0$.
\end{Proof}

Another very popular reduction is given by
\be  \label{UJ}
      U_\mu (- \lambda) = J U_\mu (\lambda) J^{-1} \ ,   \qquad  J^2 = c_0 I \ ,
\ee
then one can prove that $\Psi ( - \lambda) = J \Psi (\lambda) C (\lambda)$,
and we choose such initial conditions that $C (\lambda) = J^{-1}$, i.e.,
\be \label{DPsiJ} 
    \Psi ( - \lambda) = J \Psi (\lambda) J^{-1} \ , \qquad
 D (- \lambda) = J D (\lambda) J^{-1} \ .
\ee
Such choice of $C (\lambda)$ is motivated by a natural requirement that $\Psi, {\tilde \Psi}, D$ 
are elements of the same loop group (by the way, the formula \rf{Psiun} has the same 
motivation).

\begin{lem}
We assume \rf{Dlamu}. Then the reduction \rf{DPsiJ} imposes the following constraints on the Darboux matrix \rf{newD}:
\be
\lambda_- = - \lambda_+ \ , \qquad \varphi_+ = \varphi_- \ , \qquad
      d_+ =  J^{-1} d_- J \ , 
\ee
and (for $n = 2$) \ $ v_- = J v_+$.
\end{lem}

\begin{Proof}
From \rf{DPsiJ} it follows that $\det D (\lambda) = \det D (-\lambda)$ which means that 
the set of roots of the equation $\det D (\lambda) = 0$ is invariant under the transformation
$\lambda \rightarrow - \lambda$. Therefore $\lambda_- = - \lambda_+$. Then, using once 
more \rf{DPsiJ} we get $D_- = J D_+ J^{-1}$ and $\Psi_- = J \Psi_+ J^{-1}$. Hence
$ \varphi_+ d_+ = \varphi_- J^{-1} d_- J $. Thus $\varphi_+ = c_0 \varphi_-$, where $c_0$ is
a constant. Without loss of the generality we can take $c_0 = 1$ 
(redefining $d_\pm$ if necessary). In the case $n=2$ the kernels of $d_\pm$ are 1-dimensional.
Therefore $0 = d_+ v_+ = J^{-1} d_- J v_+ $ implies $v_- = c_1 J v_+$, where $c_1 = \const$. 
We can take  $v_+ = J v_-$.
\end{Proof}

Other types of reductions (compare \cite{Mi,Ci-dbt})
can be treated in a similar way. 

\section{The multi-soliton Darboux matrix}

In this section we generalize the approach of \cite{BC}. First, we relax the 
assumption \rf{kwadrat}. Second, we consider the $N$-soliton case 
(the Darboux matrix is a polynom of order $N$):
\be  \label{DAN}
  D (\lambda) = A_0 + A_1 \lambda + \ldots A_N \lambda^N \ .
\ee
The condition \rf{kwadrat} will be replaced by:
\be  \label{DkTk}
  D (\lambda_k) T (\lambda_k) = 0 
\ee
We denote $D_k \equiv D (\lambda_k)$, $T_k \equiv T (\lambda_k)$, 
$\Psi_k \equiv \Psi (\lambda_k)$ 
and $U_{k\mu} \equiv U_\mu (\lambda_k)$.
Evaluating  \rf{Dmu} at $\lambda = \lambda_k$ and multiplying the 
resulting equation by $T_k$ from the right we get:
\be  \label{DTk}
 D_k,_\mu T_k + D_k U_{k\mu} T_k = 0 
\ee

To solve the equation \rf{DTk}
we define $d_k$ and $h_k$ by
\be  \label{DdTc}
   D_k = \Psi_k d_k \Psi_k^{-1} \ , \qquad T_k = \Psi_k h_k \Psi_k^{-1}
\ee

\[   
D_k,_\mu = 
\Psi_k,_\mu d_k \Psi_k^{-1} + \Psi_k d_k,_\mu \Psi_k^{-1} - 
\Psi_k d_k \Psi_k^{-1} \Psi_k,_\mu \Psi_k^{-1}  \ .
\]
Therefore
\[
D_k,_\mu = U_{k\mu} D_k + \Psi_k d_k,_\mu \Psi_k^{-1} - D_k U_{k\mu} \ , 
 \]
and, taking into account \rf{DkTk} and \rf{DdTc}, we rewrite \rf{DTk} as follows
\be  \label{DTk'}
\Psi_k d_k,_\mu h_k \Psi_k^{-1} = 0 \ .
\ee
Finally, as a straightforward consequence of \rf{DkTk} and 
\rf{DTk'} we get the following constraints 
on $d_k$ and $h_k$:
\be  \label{dcmu}
     d_k h_k = 0 \ , \qquad d_k h_k,_\mu = 0 \ .
\ee
In \cite{BC} we confined ourselves to the case $T(\lambda) = D(\lambda)$, i.e, 
$d_k = \varphi_k d_{0k}$, 
($\varphi_k$ scalar functions, $d_{0k}$ constant elements satisfying $d_{0k}^2 = 0$), $h_k = d_k$. Now we are going to obtain the general solution of
\rf{dcmu} in the case of $2 \times 2$ matrices.

\begin{lem}
Let $d$ and $h$ are $2\times 2$ matrices depending on $x^1,\ldots,x^n$
such that $d h = 0$,  $d h,_\mu = 0$ and $d\neq 0$, $h \neq 0$. 
Then there exist  constants $c^1, c^2$ and  
scalar functions $q^1, q^2, p^1, p^2$ (depending on $x^1,\ldots,x^n$)
such that
\be \ba{l}   \label{dh}
 d = \mm   q^1  c^2  & - q^1 c^1  \\ 
     q^2 c^2 & - q^2 c^1  \ema = 
    \m  q^1 \\ q^2 \ema  \mm c^2 & - c^1 \ema \equiv 
 q c^\perp  ,  \\[3ex]
 h = \mm  c^1 p^1  &  c^1  p^2  \\
   c^2 p^1    &   c^2  p^2  \ema
= \m  c^1  \\  c^2 \ema  \mm  p^1 & p^2 \ema 
\equiv  c p^T 
\ea \ee
\end{lem}
\begin{Proof}
The columns of $h$ are orthogonal to the rows of $d$. If $\det (d) \neq 0$,
then, obviously, $h = 0$ in contrary to our assumptions. Therefore $\det (d) = 0$ which means that the rows of $d$ are linearly dependent. Similarly, the columns of 
$h$ are linearly dependent as well. We denote them by $p^1 c$ and $p^2 c$ (where $c$ 
is a column vector). Thus $h = c p^T$, where $p^T := (p^1, p^2)$. 

$d h = 0$ means that the columns of $h$ are
orthogonal to the rows of $d$. Therefore these rows are of the form $q^1 c^\perp$, 
$q^2 c^\perp$, where $c^\perp$ is a vector orthogonal to $c$, and, finally 
$d = q c^\perp$. Thus we obtained \rf{dh}. 

 Taking into account the condition $d h,_\mu =0$ we get
\[
  0 = q c^\perp (c,_\mu p^T + c p^T,_\mu ) = q c^\perp c,_\mu p^T 
\quad \Rightarrow \quad c^\perp c,_\mu = 0 
\]   
This means that $c^2 c^1,_\mu = c^1 c^2,_\mu$, or $c^2/c^1$ is a constant. In other 
words, $c^1 = f c^{10}$, $c^2 = f c^{20}$ ($f$ is a function, $c^{10}$, $c^{20}$ are 
constants. To complete the proof we redefine $p \rightarrow f p$, $q \rightarrow 
f q$, and $c^{k0} \rightarrow c^k$. 
\end{Proof}

Therefore, 
\be  \label{Dlk}
  D(\lambda_k) = \Psi (\lambda_k) q_k c_k^\perp \Psi^{-1} (\lambda_k) \ ,
\ee
where $c_k$ are given constant column unit vectors, $c_k^\perp$ is 
a row vector orthogonal to $c_k$ and $q_k$ are some 
vector-valued functions (column vectors). We keep the notation
$ q_k c_k^\perp \equiv d_k$, but now in general $d_k^2 \neq 0$. 

We notice that the freedom concerning the choice of $q_k$ corresponds to 
the arbitrariness of the normalization matrix. In particular, the condition
\rf{kwadrat} imposes strong constraints on ${\cal N}$. The condition 
\rf{kwadrat} can be rewritten as $q_k = \varphi_k c_k$

The constraint \rf{DkTk} implies $\det D (\lambda_k) = 0$. 
In the case of $2\times 2$ matrices the equation 
$\det D (\lambda) = 0$ (where $D$ is given by \rf{DAN}) has $2 N$ roots (at most): $\lambda_1$, \ldots
$\lambda_{2N}$.

Taking any $N+1$ pairwise different roots (say $\lambda_1,\ldots,\lambda_{N+1}$)
and 
using Lagrange's interpolation formula for polynomials, we get the
generalization of the formula \rf{newD}:
\be  \label{Lagr}
D(\lambda) = \sum_{k=1}^{N+1} \left( \prod_{\stackrel{j=1}{j\neq k}}^{N+1}
\frac{(\lambda - \lambda_j)}{(\lambda_k - \lambda_j) } \right) \Psi(\lambda_k) 
q_k c_k^\perp \Psi^{-1}(\lambda_k)
\ .
\ee

We have also $N-1$ matrix constraints which result from evaluating the 
formula \rf{Lagr} at $\lambda_{N+2},\ldots,\lambda_{2N}$:
\be  \label{suma}
  \sum_{k=0}^{N+1} \frac{ \Psi(\lambda_k) 
q_k c_k^\perp \Psi^{-1}(\lambda_k)}{ (\lambda_k - \lambda_0) \ldots 
(\lambda_k - \lambda_{k-1} ) (\lambda_k - \lambda_{k+1}) 
\ldots (\lambda_k-\lambda_{N+1}) } = 0 \ , 
\ee
where $\lambda_0 = \lambda_{N+2},\ldots,\lambda_{2N}$.

We denote
\be  \label{QkCk}
Q_k := \Psi (\lambda_k) q_k \ , \qquad C^\perp_k := c^\perp_k \Psi^{-1} (\lambda_k)
\ee

The Darboux matrix is parameterized by $2 N$ constants $\lambda_k$, $2 N$ vector functions $q_k$ and $2 N$ constant vectors $c_k$ subject to the constraints \rf{suma}. 
  
The crucial point consists in solving the system \rf{suma} in order to get 
parameterization of the Darboux matrix by a set of independent quantities. 
We plan to express $2 N -2$ functions from among $Q_1,\ldots,Q_{2N}$ by
other data. For instance, we choose $Q_1, Q_2$ as independent 
functions (they correspond to the normalization matrix ${\cal N}$). 

We rewrite the system \rf{suma} as
\be  \label{suma1}
   \sigma_{\nu 0} Q_\nu C_\nu^\perp + \sum_{k=1}^{N+1}  
 \sigma_{\nu k} Q_k C_k^\perp = 0 \ ,  \qquad (\nu = N+2,\ldots, 2N) \ ,
\ee
where
\[  \ba{l} \displaystyle
\sigma_{\nu k} = \frac{ 1 }{ (\lambda_k - \lambda_\nu) 
(\lambda_k - \lambda_1)  \ldots 
(\lambda_k - \lambda_{k-1} ) (\lambda_k - \lambda_{k+1}) 
\ldots (\lambda_k-\lambda_{N+1}) }  \\[3ex]
\displaystyle
\sigma_{\nu 0} = \frac{ 1 }{ (\lambda_\nu - \lambda_1) \ldots 
(\lambda_\nu - \lambda_N ) (\lambda_\nu - \lambda_{N+1}) } \ . 
\ea \]
Thus we have a system \rf{suma1} linear with respect to $Q_k$. 
We are going to express 
$2 N - 2$ vector functions $Q_3,\ldots,Q_{2N}$ by $Q_1, Q_2$ and the other 
parameters: $C_k, \lambda_k$. Then, using \rf{QkCk}, we could 
get $q_3,\ldots,q_{2N}$, etc. However, it is better to write \rf{Lagr}
in terms of $Q_k$:
\be  \label{Lagr1}
D(\lambda) = \sum_{k=1}^{N+1} \left( \prod_{\stackrel{j=1}{j\neq k}}^{N+1}
\frac{(\lambda - \lambda_j)}{(\lambda_k - \lambda_j) } \right) Q_k  
c_k^\perp \Psi^{-1}(\lambda_k)
\ .
\ee

Taking the scalar product of \rf{suma1} by $C_1$ we get
\be  \label{suman}
   Q_\nu  = -  \sum_{k=2}^{N+1}  
 \frac{\sigma_{\nu k} \scal{C_k^\perp}{C_1} }{ \sigma_{\nu 0}  
\scal{C_\nu^\perp}{C_1}} \ Q_k  \ ,  \qquad (\nu = N+2,\ldots, 2N) \ ,
\ee
and the scalar product of $\nu$th equation of \rf{suma1} by $C_\mu$ yields 
\be  \label{suma2}
 \sum_{k=1}^{N+1}  
 \sigma_{\nu k} \scal{C_k^\perp}{C_\nu} Q_k  = 0 \ ,  \qquad (\nu = N+2,\ldots, 2N) \ .
\ee
This is a system of $N-1$ linear equations with respect to 
$Q_1,\ldots, Q_{N+1}$. Therefore, we can (for instance) express
$Q_3,\ldots,Q_{N+1}$ in terms of $Q _1, Q_2$. Then, using \rf{suman}, 
we have $Q_{N+2},\ldots,Q_{2N}$ expressed in the similar way.

Our method is closely related to the  Neugebauer-Meinel approach \cite{NM}. Let 
$D$ is given by \rf{DAN}. 
We denote by $F (D (\lambda))$ the adjugate (or adjoint) 
matrix of $D$ which is, obviously, a polynom in $\lambda$. Thus
\be
      D(\lambda) F ( D(\lambda)) = w(\lambda) {I } 
\ee
where $w(\lambda) = \det (D (\lambda))$ is a scalar polynom and ${I }$ is the identity matrix.

Therefore, we can put $T (\lambda) = F ( D (\lambda))$
in the formula \rf{DkTk}
and identify $\lambda_k$ with zeros of $\det D (\lambda)$.

In the Neugebauer approach the matrix coefficients $A_k$ of the Darboux matrix are
obtained by solving the following system 
\be  \label{NM}
D (\lambda_k) \Psi (\lambda_k) c_k = 0  \ , (k=1,\ldots,nN)
\ee
where $\lambda_k$ and constant vectors $c_k$ are treated as given parameters.
Thus one has $n^2 N$ scalar equations for $(N+1)n^2$ scalar variables. One
of the matrices $A_k$, say $A_N$, is considered as undetermined normalization matrix.

We point out that $D (\lambda_k)$ given by the formula 
\rf{Dlk}  satisfy \rf{NM}.

\section{The discrete case}

The discrete analogue of \rf{problin} is the
following system of linear difference equation
\be
  T_\mu \Psi = U_\mu \Psi \ ,  
\qquad (\mu=1,\ldots,m) \ ,
\ee
where $T_\nu$ denotes the shift in $\nu$th variable, 
i.e., $(T_\nu \Psi) (x^1,\ldots,x^{\nu},\ldots,x^m) := 
\Psi (x^1,\ldots,x^{\nu} + 1,\ldots,x^m)$.
The Darboux transformation is defined in the standard way:
\be
{\tilde \Psi} = D \Psi \ , \qquad 
T_\mu {\tilde \Psi} = {\tilde U}_\mu {\tilde \Psi} \ .
\ee
Therefore  $(T_\mu D) (T_\mu \Psi) = {\tilde U}_\mu D \Psi$,
and, finally
\be  \label{DDT}
    (T_\mu D) U_\mu = {\tilde U}_\mu  D 
\ee
If $D^2 (\lambda_1) = 0$, then multiplying \rf{DDT} by $D (\lambda)$ from the right, and
evaluating the obtained equation at $\lambda=\lambda_1$ we see that the right hand side 
vanishes and we get:
\be  \label{eqD1}
       (T_\mu D_1) U_\mu (\lambda_1) D_1 = 0
\ee
where we denote $D_1 := D(\lambda_1)$. In order to solve \rf{eqD1} we put
\[
D_1 = \varphi_1 \Psi_1 d_1 \Psi_1^{-1}
\]
where $\Psi_1 := \Psi (\lambda_1)$. Then \rf{eqD1} takes the form:
\[
 \varphi_1  T_\mu (\varphi_1) (T_\mu \Psi_1) (T_\mu d_1) d_1 \Psi_1^{-1} = 0 \ .
\]
Therefore, if 
\be  \label{Tdd}
    (T_\mu d_1 ) d_1 = 0
\ee
then the equation \rf{eqD1} is satisfied.
The condition \rf{Tdd} can be rewritten (at least in the matrix case) as
\[
  {\rm Im} d_1 \subset {\rm ker} (T_\mu d_1) 
\]
In other words, the sequence of linear operators
\[
  \ldots \ \rightarrow \quad T_\mu^{-1} d_1 \ \rightarrow \quad d_1 \ \rightarrow
\quad T_\mu d_1 \ \rightarrow \quad T_\mu^2 d_1 \ \rightarrow \  \ldots 
\]
is an exact sequence \cite{Lang}. 

Similarly as in the smooth case we mostly confine ourselves to the simplest 
solution of \rf{Tdd}, i.e., $d_1 = \const$ which implies  $d_1^2 = 0$.
The Darboux matrix has the same form  \rf{newD} as in the continuum case.

{ \bf  Summary.}  
In this paper we developed 
the approach of \cite{BC} considering explicitly the most 
important reductions, extending our results on the  
$N$-soliton case,  and showing that the discrete case is,
as usual, very similar to the continuous one.


\begin{thebibliography}{99}

\newcommand{\vbib}{\par \vspace{-2ex} \par \bib}

\footnotesize

\par \vspace{-2ex} \par \bibitem{ZS}
      V.E.Zakharov, A.B.Shabat: 
``Integration of nonlinear equa\-tions of mathematical physics by the
inverse scattering met\-hod.\  \-II'',
{\it Funk.\  \-Anal.\  Pril.} {\bf 13} (1979), 13-22 [in Russian].



\par \vspace{-2ex} \par \bibitem{Mi}
      A.V.Mikhailov: 
``The reduction problem and the inverse scattering me\-t\-hod'',
           {\it Physica} {\bf D 3} (1981), 73-117.

\par \vspace{-2ex} \par \bibitem{NM}
                G.Neugebauer, R.Meinel: 
``General $N$-soliton solution of the AKNS class on arbitrary background'',
          {\it Phys.\  Lett.} {\bf A 100} (1984), 467-470.
             
\par \vspace{-2ex} \par \bibitem{Its}
       A.R.Its:
    ``Liouville's theorem and the inverse scattering method'', 
  [in:] {\it Zapiski Nau.\ Sem.\ LOMI} {\bf 133} (1984), 113-125
 [in Russian].


\par \vspace{-2ex} \par \bibitem{MS}
      V.B.Matveev,  M.A.Salle: {\it    Darboux  Transformations   and 
     Solitons}, Sprin\-ger-Verlag, Berlin-Heidelberg 1991.

\par \vspace{-2ex} \par \bibitem{Gu}
    C.H.Gu:
      ``B\"acklund Transformations and Darboux Transformations'',
         [in:] {\it Soliton Theory and Its Applications}, 
                Springer, Berlin 1995, pp.\ 122-151.     

\par \vspace{-2ex} \par \bibitem{Ci-dbt}
                J.Cie\'sli\'nski: 
                ``An algebraic method to construct the Darboux matrix'', 
                {\it J.Math.Phys.} {\bf 36} (1995) 5670-5706.

\par \vspace{-2ex} \par \bibitem{RS}
C.Rogers, W.K.Schief:
{\it B\"acklund and Darboux transformations. Geometry and modern applications 
in soliton theory}, Cambridge Univ.\ Press, Cambridge 2002.


\par \vspace{-2ex} \par \bibitem{CGS-izot}
J.Cie\'sli\'nski, P.Goldstein, A.Sym: 
``Isothermic surfaces in $E^3$ as soliton surfaces'', 
{\it Phys. Lett.} {\bf A 205} (1995) 37-43.


\par \vspace{-2ex} \par \bibitem{Ci-izot}
J.Cie\'sli\'nski: 
``The Darboux-Bianchi transformation for isothermic surfaces. Classical
results versus the soliton approach'', {\it Diff.\ Geom.\ Appl.}
{\bf 7} (1997), 1-28.


\par \vspace{-2ex} \par \bibitem{Ci-Cliff}
J.L.Cie\'sli\'nski:
``A class of linear spectral problems in Clifford algebras'',
{\it Phys.\ Lett.\ } {\bf A 267} (2000) 251-255.



\par \vspace{-2ex} \par \bibitem{Ci-TMP}
 J.L.Cie\'sli\'nski:
``Geometry of submanifolds derived from Spin-valued spectral 
problems'', {\it Theor.\ Math.\ Phys.} {\bf 137} (2003)
1396-1405.



\par \vspace{-2ex} \par \bibitem{BC}
W.Biernacki, J.L.Cie\'sli\'nski:
``A compact formula for the Darboux-B\"acklund transformation for some 
spectral problems in Clifford algebras'',
{\it Phys.\ Lett.\ } {\bf A 288} (2001) 167-172.


\par \vspace{-2ex} \par \bibitem{ZMNP}
   V.E.Zakharov, S.V.Manakov, S.P.Novikov, L.P.Pitaievsky:  
{\it Theory of so\-li\-tons}, Nauka, Moscow 1980 [in Russian].

\par \vspace{-2ex} \par \bibitem{MNS}
      R.Meinel, G.Neugebauer, H.Steudel: {\it   Solitonen. Nichtlineare
                Strukturen},   Academie Verlag, Berlin 1991 [in German].



\par \vspace{-2ex} \par \bibitem{Ci-nos}
J.L.Cie\'sli\'nski:
{\it The Darboux-Bianchi-B\"acklund transformation and soliton surfaces},
[in:] {\it Nonlinearity and Geometry} (edited by D.W\'ojcik
and J.L.Cie\'sli\'nski), Polish Scientific Publishers PWN, Warsaw 1998.



\par \vspace{-2ex} \par \bibitem{Lang}
 S.Lang:
{\it Algebra}, Addison-Wesley Publ.\ Co.\ 1965.


\end{thebibliography}
\end{document}